\documentclass[12pt]{article}
\usepackage[hmargin=2.5cm,vmargin=2cm]{geometry}
\usepackage{graphicx}
\usepackage{amsmath}
\usepackage{enumerate}

\begin{document}
\title{Real-time particle-detection probabilities in accelerated macroscopic detectors }
\author {Charis Anastopoulos\footnote{anastop@physics.upatras.gr} and   Ntina Savvidou\footnote{ksavvidou@physics.upatras.gr}\\
 {\small Department of Physics, University of Patras, 26500 Greece} }

\maketitle

\begin{abstract}
We construct the detection rate for particle detectors moving along non-inertial trajectories and interacting with quantum fields.  The detectors described here are characterized by the presence of records of observation throughout their history, so that the detection rate corresponds to directly measurable quantities. This is in contrast to past treatments of detectors, which actually refer to probes, i.e.,
  microscopic systems from which we extract information {\em only  after} their interaction has been completed. Our treatment incorporates the irreversibility due to the creation of macroscopic records of observation. The key result is a {\em real-time}  description of particle detection and a rigorously defined time-local probability density function (PDF). The PDF depends on the scale $\sigma$ of the temporal coarse-graining that is necessary for the formation of a macroscopic record.  The evaluation of the PDF for Unruh-DeWitt detectors along different types of trajectory shows that only paths with at least one characteristic time-scale much smaller than $\sigma$ lead to appreciable particle detection. Our approach allows for averaging over fast motions and thus predicts a constant detection rate for all fast periodic motions.

\end{abstract}

\section{Introduction}

A particle detector moving along a non-inertial trajectory in Minkowski spacetime and interacting with a quantum field records particles, even if the quantum field lies in the vacuum state \cite{Unruh, Boyer, Dewitt}. For trajectories with constant proper acceleration $a$, the detected particles are distributed according to a Planckian spectrum, at the Unruh temperature $T = \frac{a}{2\pi}$. However, the thermal response of the accelerated detectors is only local. As shown in Ref. \cite{AnSav11},  a spatially separated pair of particle detectors with the same proper acceleration $a$ has non-thermal temporal correlations of the detection events.

In Ref. \cite{AnSav11} we established the importance of a distinction between probes and particle detectors, in any discussion of the response of the quantum vacuum to accelerated motion. In particular,
probes are microscopic systems that interact with the quantum field through their accelerating motion. After the interaction has been completed,  information can be extracted from a probe through a single measurement. The mathematical description of probes is very similar to that of particle scattering.
  In contrast, particle detectors are characterized by the macroscopic records of detection events. These macroscopic records are present {\em throughout the detectors' history}, so that  a detection rate that corresponds to directly measurable quantities can be defined. Such a description of a particle detector is compatible with the standard use of the term in physics. Our approach is distinguished from most existing studies of the issue which employ the word "detector" for systems that are best described as probes. The difference between probes and detectors is further analyzed in Sec. 2: it leads to a different mathematical description and to different expressions for the particle detection rate.

In this article, we undertake a description of particle detectors moving along general spacetime trajectories, by incorporating explicitly the creation of macroscopic records into physical description. The key result is a {\em real-time}  description of particle detection and the rigorous construction of the associated probabilities.

To this end, we employ a general method for the construction of Quantum Temporal Probabilities (QTP) for any experimental configuration, that was developed in Ref. \cite{AnSav12}.   The key  idea is to distinguish between the roles of time as a parameter to Schr\"odinger's equation and as a label of the causal ordering of events \cite{Sav99, Sav10}. This important distinction leads to the definition of quantum temporal observables. In particular, we identify the time of a detection event as a coarse-grained quasi-classical variable   associated with macroscopic records of observation. Besides the construction of particle detectors, the QTP method has been applied to the time-of-arrival problem \cite{AnSav12, AnSav06}, to the temporal characterization of tunneling processes \cite{AnSav08, AnSav13} and to non-exponential decays \cite{An08}.

In Sec. 3, we present a detailed construction of the detection probability for moving particle detectors using the QTP method. This generalizes the results of Ref. \cite{AnSav11} derived for the case of constant proper acceleration. The detection probability is sensitive to a time-scale $\sigma$ that characterizes the detector's degree of coarse-graining.That is, $\sigma$ corresponds to the minimal temporal localization in time of a detection event. The time-scale $\sigma$ is a physical parameter that is determined by a detailed knowledge of the detector's physics.

The most important feature of the derived probability density is that it is local in time for scales of observation much larger than $\sigma$. This property significantly simplifies the calculation of the detection probability for a large class of spacetime paths. Such calculations are presented in Sec. 4. The most important results are (i) the derivation of the adiabatic approximation and of its corrections and (ii) the demonstration that the   detection probability of fast periodic motions is constant.

Finally, in Sec. 5 we summarize our results, discussing in particular their relevance to proposed experiments.

\section{ Distinction between probes and detectors}

In order to explain the difference between a probe and a detector, we consider the most common model employed in this context, the Unruh-DeWitt detector \cite{Unruh, Dewitt}. This is a quantum system that moves along a trajectory $X^{\mu}(\tau)$ in Minkowski spacetime and that is characterized by a Hamiltonian $\hat{H}_0$. The Unruh-DeWitt detector is coupled to a massless, free quantum scalar field $\hat{\phi}(x)$ through an interaction Hamiltonian $\hat{H}_{int} =   \hat{m} \otimes \hat{\phi}[X(\tau)]$, where $\hat{m}$ is an operator analogue of the dipole moment.

Assuming that the detector is initially at the lowest energy state $| 0 \rangle$ and the field is at the vacuum state $|\Omega \rangle$, the  probability $\mbox{Prob}(E, \tau)$ that  a measurement of the detector will find   energy $E > 0$ {\em at proper time $\tau$} is
\begin{eqnarray}
\mbox{Prob}(E, \tau) = Tr \left[(\hat{P}_E \otimes 1) (T e^{ -i \int_{\tau_0}^{\tau} ds\hat{H}_{int}(s)  }) \left(|0 \rangle \langle 0| \otimes |\Omega \rangle \langle \Omega|\right)  (T e^{ -i \int_{\tau_0}^{\tau} ds\hat{H}_{int}(s)  })^{\dagger}\right], \label{refbad}
\end{eqnarray}
where $\hat{P}_E$ is the projector onto the eigenstates of the detector with energy equal to $E$ and $T e^{ -i \int_{\tau_0}^{\tau} ds\hat{H}_{int}(s)  }$ is the time-ordered product.

To first order in perturbation theory, $Te^{- i \int_{\tau_0}^{\tau} ds\hat{H}_{int}(s)  } = 1 -  i \int_{\tau_0}^{\tau} ds\hat{H}_{int}(s) $ and Eq. (\ref{refbad}) becomes
\begin{eqnarray}
\mbox{Prob}(E, \tau) &=& \alpha(E) \int_{\tau_0}^{\tau} ds  \int_{\tau_0}^{\tau} ds' e^{-iE(s'-s)} W[X(s'), X(s)] \nonumber
\\
&=& 2 \alpha(E) \mbox{Re} \int_{\tau_0}^{\tau} ds' \int_0^{s' - \tau_0} ds e^{-iEs}  W[X(s'), X(s' - s)]  ,\label{refbad2}
\end{eqnarray}
where $\alpha(E) = \langle 0|\hat{m} \hat{P}_E \hat{m}|0 \rangle $ and $W(X',X) = \langle \Omega|\hat{\phi}(X')\hat{\phi}(X)|\Omega\rangle$ is the field's Wightman's function. In the derivation of Eq. (\ref{refbad2}) one assumes that the detector-field interaction is  switched on only  for a finite time interval of duration $T$ \cite{SvSv, Hig93, SPad96}.

The time derivative of $\mbox{Prob}(E, \tau)$ is often identified with the transition probability $P(E, \tau)$, i.e. with a Probability Density Function (PDF) such that $P(E, \tau) \delta t$ equals with the probability that a transition to energy $E$ occurred at some time within the interval $[\tau,  \tau + \delta \tau]$. Hence, one often defines
\begin{eqnarray}
P(E, \tau) := \frac{d}{d \tau} Prob(E, \tau) =  2 \alpha(E) \mbox{Re}   \int_0^{\tau - \tau_0} ds e^{-iEs}  W(s', s' - s) \label{petbad} .
\end{eqnarray}
With a suitable regularization of the Wightman's function, Eq. (\ref{petbad}) can be made fully causal \cite{Schlicht, Langlois, LoukoSatz, Milgrom}, i.e.,  variables at time $\tau$ depend only on the trajectory at times prior to $\tau$. However, Eq. (\ref{petbad}) is highly non-local in time; one needs to know the full past history of the detector in order to identify the transition probability at the present moment of time.

There are two problems with the line of reasoning that leads to the consideration of Eq. (\ref{petbad}) as a  PDF for the detection rate. First, the above definition of a PDF is not  probabilistically sound, because the derivative of $\mbox{Prob}(E, \tau)$ with respect to $\tau$ does not, in general, define a probability measure. The time $\tau$ in $Prob(E, \tau)$ is not a random variable (unlike $E$), but a {\em parameter} of the probability distribution. Thus, the time derivative of $Prob(E, \tau)$ does not define a PDF; in general, it takes negative values. The fact that $P(E, \tau)$ in Eq. (\ref{petbad}) is positive-valued is an artifact of first-order perturbation theory. Second order effects include the relaxation of the detector through particle emission, which render $\mbox{Prob}(E, \tau)$  a decreasing function of $\tau$ at later times \cite{HuLin}. Thus, negative values of $P(E, \tau)$ appear.

Second,  Eq. (\ref{refbad}) applies to experimental configurations at which information is extracted from  the Unruh-DeWitt detector   only once, at time $\tau$.   Hence,   Eq. (\ref{petbad}) does not describe a system that actually  {\em records}  particles   at times prior to $\tau$; it purports to define a detection rate for fictitious   detection events. In fact, the physical system described by Eq. (\ref{refbad}) cannot be described as a detector, in any reasonable use of the word in physics. We usually think of a particle detector (e.g. a photodetector) as a physical system that outputs a time-series of detection events, each event being well localized in time. The quantum mechanical description of the detector ought to take into account the {\em irreversible} process of outputting information, namely, the creation of macroscopic records of observation.

However, the system described by Eq. (\ref{refbad}) outputs information only once, at time $\tau$, and its time evolution prior to $\tau$ is fully unitary. Such a system is best described as a field {\em probe}, in the sense of Bohr and Rosenfeld \cite{BoRo}: a probe is a microscopic system that interacts with the quantum field; information about the field in incorporated in the probe's final state and we extract this information through a single measurement.
  For such a system, Eq. (\ref{refbad}) provides a good approximation for times much earlier than the system's relaxation time $\Gamma^{-1}$.

Another key difference between detectors and probes is that in the former it is possible to determine  {\em temporal} correlation functions for particle detection, by considering several detectors along different spacetime trajectories \cite{AnSav11}. Measurements of such correlations (for static detectors) are well established in quantum optics \cite{WM}.

Next, we proceed to the construction of models for particle detectors characterized by macroscopic records of observation.

\section{Macroscopic particle detectors}
In this section, we derive   the detection rate of an Unruh-Dewitt detector moving along a general trajectory in Minkowski spacetime. First, we briefly review the QTP method for defining PDFs with respect to detection time. Then, we construct explicitly a model for a macroscopic Unruh-Dewitt detector.

\subsection{A review of the QTP method}

 We follow the general methodology for constructing detection probabilities with respect to time, that was developed in Ref. \cite{AnSav12} (the Quantum Temporal Probabilities method). The reader is referred to this article for a detailed presentation.
The key result of Ref. \cite{AnSav12} is the derivation of a general formula for probabilities associated with the time of an event  in a general quantum system. Here,   the word "event" refers to a definite and persistent macroscopic record of observation. The event time $t$ is a coarse-grained, quasi-classical parameter associated with such records: it corresponds to the reading of an external classical clock that is simultaneous with the emergence of the record.

Let  ${\cal H}$ be the
Hilbert space associated with  the physical   system under consideration;   ${\cal H}$ describes the degrees of freedom of    microscopic particles and of a macroscopic measurement apparatus.
 We assume that
 ${\cal H}$ splits into two subspaces: ${\cal H} = {\cal
H}_+ \oplus {\cal H}_-$. The subspace ${\cal H}_+$ describes the accessible
states of the system given that a specific event is realized; the subspace ${\cal H}_-$ is the complement of ${\cal H}_+$. For example, if the quantum event under consideration is a detection of a particle by a macroscopic apparatus, the subspace ${\cal H}_+$ corresponds to all accessible states of the apparatus given that a detection event has been recorded. We denote  the
projection operator onto ${\cal H}_+$ as $\hat{P}$ and
the projector onto ${\cal H}_-$ as $\hat{Q} := 1  - \hat{P}$.

 We note that the transitions under consideration are always correlated with the emergence of a macroscopic observable    that is recorded as a measurement outcome.    In this sense, the transitions considered here are {\em irreversible}. Once they occur, and a measurement outcome has been recorded, the further time evolution of the degrees of freedom in the measurement device is irrelevant to the probability of transition.

Once a transition has taken place,   the values of a microscopic variable are determined through correlations with a pointer variable of the measurement apparatus. We denote by $\hat{P}_\lambda$   projection operators (or, more generally, positive operators) that correspond to different values $\lambda$ of some physical magnitude. This physical magnitude can be measured only if the quantum event under consideration has occurred. For example, when considering transitions associated with particle detection, the projectors $\hat{P}_\lambda$  may be correlated  to properties of the microscopic particle, such as position, momentum and energy.
The set of projectors $\hat{P}_\lambda$ is exclusive ($\hat{P}_{\lambda} \hat{P}_{\lambda'} = 0, $ if $\lambda \neq \lambda'$). It is also exhaustive given that the event under consideration  has occurred; i.e., $\sum_\lambda \hat{P}_\lambda = \hat{P}$.

We also assume that  the system is initially ($t = 0$) prepared at a
state $|\psi_0 \rangle \in {\cal H}_-$, and that   time evolution is
governed by the self-adjoint Hamiltonian operator $\hat{H}$.

In Ref. \cite{AnSav12}, we derived
 the probability amplitude $| \psi; \lambda, [t_1, t_2] \rangle$ that corresponds to (i) an initial state $|\psi_0\rangle$, (ii) a transition occurring at some instant in the time interval $[t_1, t_2]$ and (iii) a recorded value $\lambda$ for the measured observable:

 \begin{eqnarray}
| \psi_0; \lambda, [t_1, t_2] \rangle = - i e^{- i \hat{H}T}
\int_{t_1}^{t_2} d t \hat{C}(\lambda, t) |\psi_0 \rangle.
\label{ampl5}
\end{eqnarray}
where   the {\em class operator} $\hat{C}(\lambda, t)$ is defined as
\begin{eqnarray}
\hat{C}(\lambda, t) = e^{i \hat{H}t} \hat{P}_{\lambda} \hat{H}
\hat{S}_t, \label{class}
\end{eqnarray}
and $\hat{S}_t =  \lim_{N \rightarrow \infty}
(\hat{Q}e^{-i\hat{H} t/N} \hat{Q})^N$
is the restriction of the propagator in ${\cal H}_-$. The parameter $T$ in Eq. (\ref{ampl5}) is a reference time-scale at which the amplitude is evaluated. It defines an upper limit to $t$ and it corresponds to the duration of an experiment. It cancels out when evaluating probabilities, so it does not appear in the physical predictions.

If $[\hat{P}, \hat{H}] = 0$, i.e., if the Hamiltonian
evolution preserves the subspaces ${\cal H}_{\pm}$, then $|\psi_0;
\lambda, t \rangle = 0$. For a Hamiltonian    of the form $\hat{H} = \hat{H_0} + \hat{H_I}$, where $[\hat{H}_0, \hat{P}] = 0$, and $H_I$ a perturbing interaction, we obtain
\begin{eqnarray}
\hat{C}(\lambda, t) = e^{i \hat{H}_0t} \hat{P}_{\lambda} \hat{H}_I
e^{-i \hat{H}_0t}, \label{perturbed}
\end{eqnarray}
 to leading order in the perturbation.

The benefit of Eq. (\ref{perturbed}) is that it does not involve the restricted propagator $\hat{S}_t$, which is difficult to compute.

The amplitude  Eq. (\ref{ampl5}) squared defines  the probability $\mbox{Prob} (\lambda, [t_1, t_2])\/$
that at some time in the interval $[t_1, t_2]$ a detection with
outcome $\lambda$ occurred
\begin{eqnarray}
\mbox{Prob}(\lambda, [t_1, t_2]) := \langle \psi_0; \lambda, [t_1, t_2] | \psi_0;
\lambda, [t_1, t_2] \rangle =   \int_{t_1}^{t_2} \,  dt
\, \int_{t_1}^{t_2} dt' \; Tr [\hat{C}(\lambda, t) \hat{\rho}_0
\hat{C}^{\dagger}(\lambda, t) ], \label{prob1}
\end{eqnarray}
where $\hat{\rho}_0 = |\psi_0\rangle \langle \psi_0|$.

However,   $\mbox{Prob}(\lambda, [t_1, t_2])$ does not correspond to a well-defined probability measure because it
fails to satisfy the Kolmogorov additivity condition for probability measures
\begin{eqnarray}
\mbox{Prob}(\lambda, [t_1, t_3]) = \mbox{Prob}(\lambda, [t_1, t_2]) + \mbox{Prob}(\lambda, [t_2,
t_3]).  \label{kolmogorov}
\end{eqnarray}

Eq. (\ref{kolmogorov}) does not hold for generic choices of $t_1, t_2$ and $t_3$.   The key point is that in a macroscopic system (or in a system with a macroscopic component) one expects that Eq. (\ref{kolmogorov}) holds with a good degree of approximation, given a sufficient degree of coarse-graining \cite{Omn, Omn2, Gri, GeHa93, hartlelo}. Thus, if the time of transition is associated with a macroscopic measurement record, there exists a coarse-graining time-scale $\sigma$, such that    Eq. (\ref{kolmogorov}) holds, for $ |t_2 - t_1| >> \sigma$ and $|t_3 - t_2| >> \sigma$. Then, Eq. (\ref{prob1}) does define a probability measure when restricted to intervals of size  larger than $\sigma$.

 It is convenient to proceed by smearing the amplitudes
 Eq. (\ref{ampl5}) at a time-scale of order $\sigma$. To this end, we introduce a family of probability density functions $f_{\sigma}(s)$,  localized around $s = 0$ with width $\sigma$, and normalized so that
$\lim_{\sigma \rightarrow 0} f_{\sigma}(s) = \delta(s)$. The only requirement  is that $f_{\sigma}$ satisfies approximately the equality

\begin{eqnarray}
\sqrt{f_{\sigma}(t-s) f_{\sigma}(t-s')} = f_{\sigma}(t - \frac{s+s'}{2}) g_{\sigma}(s-s'), \label{eq2}
\end{eqnarray}
where $w_{\sigma}(s)$ is a function strongly localized around $s = 0$. From Eq. (\ref{eq2}), it trivially follows that  $w_{\sigma}(0) = 1$. As an example, we note that   the Gaussians
$f_{\sigma}(s) = (2 \pi \sigma^2)^{-1/2}
\exp\left[-\frac{s^2}{2\sigma^2}\right]$
satisfy Eq. (\ref{eq2}) {\em exactly}, with $w_{\sigma}(s) = \exp[-s^2/(8\sigma^2)]$.

We  define the smeared amplitude $|\psi_0; \lambda, t\rangle_{\sigma}$   as
\begin{eqnarray}
|\psi_0; \lambda, t\rangle_{\sigma} := \int ds \sqrt{f_{\sigma}(s -t)}
|\psi_0; \lambda, s \rangle = \hat{C}_{\sigma}(\lambda, t) |\psi_0
\rangle, \label{smearing}
\end{eqnarray}
where
\begin{eqnarray}
\hat{C}_{\sigma}(\lambda, t) := \int ds \sqrt{f_{\sigma}(t - s)}
\hat{C}(\lambda, s).
\end{eqnarray}
The square amplitudes
\begin{eqnarray}
\bar{P}(\lambda, t) = {}_{\sigma}\langle \psi_0; \lambda, t|\psi_0;
\lambda, t\rangle_{\sigma} = Tr \left[\hat{C}^{\dagger}_{\sigma}(\lambda, t)
 \hat{\rho}_0 \hat{C}_{\sigma}(\lambda, t)\right] \label{ampl6}
\end{eqnarray}
define a PDF with respect to the time $t$ that corresponds to a Positive-Operator-Valued-Measure.

Using Eq. (\ref{eq2}), the PDF Eq. (\ref{ampl6}) becomes
\begin{eqnarray}
\bar{P}(\lambda,t)  = \int ds f_{\sigma}(t - s) P(\lambda, s), \label{conv}
\end{eqnarray}
where
\begin{eqnarray}
P(\lambda, t) = \int ds w_{\sigma}(s)  Tr \left[\hat{C}^{\dagger}_{\sigma}(\lambda, t+ \frac{s}{2})
 \hat{\rho}_0 \hat{C}_{\sigma}(\lambda, t - \frac{s}{2})\right]. \label{probmain}
\end{eqnarray}
This means that the PDF $\bar{P}(\lambda, s)$ is the convolution of $P(\lambda, s)$ with weight corresponding to $f_{\sigma}$. For scales of observation much larger than $\sigma$ the difference between $P(\lambda, s)$ and  $\bar{P}(\lambda, s)$ is insignificant. Hence, it is usually sufficient to work with the deconvoluted form of the probability density, Eq. (\ref{probmain}), which is simpler. An important exception is Sec. 3.5. Note also that other sources of measurement error (for example, environmental noise) lead to further smearing of the PDF Eq. (\ref{conv}). In this sense, the PDF $P(\lambda, t)$ corresponds to the ideal case that (i) any external classical noise vanishes and (ii) the scale of observation is much larger than the temporal coarse-graining scale  $\sigma$.

\subsection{Modeling an Unruh-DeWitt detector}

Next, we employ Eq. (\ref{probmain}) for constructing a PDF with respect to time for   particle detection along non-inertial trajectories.  The system under consideration consists  of a quantum field $\hat{\phi}$ that describes microscopic particles  and a macroscopic particle detector in motion.
The Hilbert space for the total system is the tensor product ${\cal F} \otimes {\cal H}_d$. The Hilbert space ${\cal H}_d$  describes the detector's degrees of freedom and the Fock space ${\cal F}$  is associated with the field $\hat{\phi}$. Here, we take $\hat{\phi}$ to be a massless scalar field in Minkowski spacetime  but the methodology can be easily adapted to more elaborate set-ups.

We assume that the detector is   sufficiently small  so  that its motion is described by a single spacetime path $X^{\mu}(\tau)$, where $\tau$ is the proper time along the path. An important assumption is that the physics at the detector's rest frame is independent of the path followed by the detector: the evolution operator for an Unruh-DeWitt detector is   $e^{-i\hat{H}_d\tau}$, where $\tau$ is the proper time along the detector's path and $\hat{H}_d$ is the Hamiltonian for a stationary detector.  Thus, path-dependence enters into the propagator only through the proper time. For a detector that follows a timelike path $X^{\mu}(\tau)$ in Minkowski spacetime, the equation $X^0(\tau) = t$ can be solved for $\tau$, in order to determine the proper time $\tau(t)$ as a function of the inertial time coordinate $t$ of Minkowski spacetime. The evolution operator for the detector with respect to the coordinate time $t$ is then
$e^{-i\hat{H}_d\tau(t)}$.

  We model  a detector that records the energy of field excitations. The relevant transitions correspond to changes in the detector's energy, as it absorbs a field excitation (particle). We assume that the initially prepared state of the detector $|\Psi_0 \rangle$ is a state of  minimum energy $E_0 = 0$, and that the excited states have energies $E > E_0$.
  We denote the projectors onto the constant energy subspaces as $\hat{P}_E$. Thus, the projectors $\hat{P}_{\pm}$ on ${\cal H}$ corresponding to the transitions are $\hat{P}_+ = \left(\sum_{E>E_0} \hat{P}_E \right) \otimes \hat{1}$ and $\hat{P}_- = \hat{P}_0 \otimes \hat{1}$, where  $\hat{P}_0 = \hat{P}_{E_0}$.

The energy of an absorbed particle is determined within an uncertainty $\Delta E << E_0$ intrinsic to the detector. This measurement is modeled by a family of   positive operators on ${\cal H}_d$
  \begin{eqnarray}
  \hat{\Pi}_E = \int d\bar{E} F_{\Delta E} (E- \bar{E}) \hat{P}_{\bar{E}},
  \end{eqnarray}
where  $F_{\Delta E}$ is a smearing function of width $\Delta E$.

The scalar field Hamiltonian is $\hat{H}_{\phi} = \int d^3 x \left( \frac{1}{2} \hat{\pi}^2 + \frac{1}{2} (\nabla \phi)^2\right)$. The evolution operator for the field with respect to a Minkowskian inertial time coordinate $t$  is $e^{- i \hat{H}_{\phi}t}$, where $t$ is a Minkowski inertial time parameter.

We also consider a local interaction Hamiltonian
\begin{eqnarray}
\hat{H}_{I} = \int dx \hat{\phi}(x) \otimes \hat{J}(x),   \label{hint}
\end{eqnarray}
where $\hat{J}({\bf x})$ is a current operator  on the Hilbert ${\cal H}_a$ of the detector.

Then Eq. (\ref{probmain}) yields
\begin{eqnarray}
P(E, \tau) = \int dy w_{\sigma}(y) \int ds ds' \int d^3x d^3x' Tr\left[ \hat{\phi}(x', s') \hat{\phi}(x, s) \hat{\rho}_0 \right]
 \delta [s - X^0(\tau + \frac{y}{2})]   \delta [s' - X^0(\tau - \frac{y}{2})] \nonumber \\
 \times \langle \Psi_0|\hat{J}(x',\tau -\frac{y}{2}) \sqrt{\hat{\Pi}_E} e^{-i \hat{H} y} \sqrt{\hat{\Pi}_E} \hat{J}(x,\tau +\frac{y}{2})|\Psi_0 \rangle, \label{pet3}
\end{eqnarray}
where $\hat{\rho}_0$ is the initial state of the field, $\hat{\phi}(x, s) = e^{i \hat{H}_0s} \hat{\phi}(x) e^{-i \hat{H}_0s}$ and $\hat{J}(x, \tau) = e^{i \hat{H}_d \tau} \hat{J}(x) e^{-i \hat{H}_d \tau}$ are the Heisenberg-picture operators for the scalar field and the current respectively.

 A defining feature of an  Unruh-Dewitt detector is that it is   point-like, i.e., effects that are related to  its finite size do not enter the detection probability. In the present context, this condition is implemented through the approximation
\begin{eqnarray}
\hat{J}(x, \tau) \simeq \hat{m} \delta^3 [x - X(\tau)], \label{UdW}
\end{eqnarray}
where $\hat{m}$ is an averaged current operator, corresponding to the analogue of the dipole moment for a scalar field.

Using Eq. (\ref{UdW}), the detection probability Eq. (\ref{pet3}) simplifies
\begin{eqnarray}
P(E, \tau) = \alpha(E) \int dy g_{\sigma}(y) e^{-iEy} W[X(\tau+\frac{y}{2}), X(\tau - \frac{y}{2})], \label{petb}
\end{eqnarray}

where
\begin{eqnarray}
\alpha(E) = Tr_{{\cal H}_d}(\hat{P}_E\hat{m} \hat{P}_0\hat{m})/Tr_{{\cal H}_d} \hat{P}_0, \label{ae}
\end{eqnarray}
 and
\begin{eqnarray}
W(X, X') = Tr\left[ \hat{\phi}(X') \hat{\phi}(X) \hat{\rho}_0 \right]
\end{eqnarray}
is the positive-frequency Wightman function. When the field is in the vacuum state,
\begin{eqnarray}
W(X, X') = \frac{-1}{4\pi^2 [(X^0-X^{0'}-i \epsilon)^2 - ({\bf X} - {\bf X'})^2]},
\end{eqnarray}
 where $\epsilon >0$ is the usual regularization parameter. In our approach the issue of carefully regularizing the Wightman function in order to preserve causality \cite{Schlicht, Tagaki} does not arise.  The detection probability is by construction local in time and causal within an accuracy defined by the temporal coarse-graining. In particular, the value of the regularization parameter $\epsilon$ is irrelevant to the physical predictions. The only use of the  term $- i \epsilon$ is to guarantee  that poles of $y$ along the real axis do not contribute to the integral  Eq. (\ref{petb}).

 Thus, the PDF for particle detection in the vacuum becomes
 \begin{eqnarray}
 P(E, \tau) = - \frac{\alpha(E)}{4 \pi^2} \int_{-i\epsilon - \infty}^{-i \epsilon + \infty} dy  \frac{ g_{\sigma}(y) e^{-i Ey}}{  \Sigma(\tau, y)}, \label{pet5}
 \end{eqnarray}
 where
 \begin{eqnarray}
 \Sigma(\tau, y) = \eta_{\mu \nu} [X^{\mu}(\tau + \frac{y}{2}) -  X^{\mu}(\tau - \frac{y}{2})] [X^{\nu}(\tau + \frac{y}{2}) -  X^{\nu}(\tau - \frac{y}{2})] \label{sigma0}
 \end{eqnarray}
 is the proper distance between two points on the detector's path characterized by values of the proper time $\tau \pm y/2$.

The function $g_{\sigma}(y)$ in Eq. (\ref{petb}) is defined as
\begin{eqnarray}
g_{\sigma}(y) = w_{\sigma}(y) \tilde{F}_{\Delta E}(y)  , \label{gs}
\end{eqnarray}
where $\tilde{F}_{\Delta E}(y) = \int dE F_{\Delta E}(E) e^{-i Ey}$. The term $\tilde{F}_{\Delta E}$ suppresses temporal interferences between amplitudes  Eq. (\ref{ampl5}) at timescales  larger than $(\Delta E)^{-1}$. Hence,  the temporal coarse-graining parameter $\sigma$ must be of order $(\Delta E)^{-1}$ or   higher for
 Eq. (\ref{kolmogorov}) to be satisfied.
 Since $\Delta E << E_0 <E$, this implies that
\begin{eqnarray}
E \sigma >> 1. \label{fund}
\end{eqnarray}
Eq. (\ref{fund})  is a defining condition for a physically meaningful  particle detector and it specifies the values of energy $E$ to which
  Eq. (\ref{petb}) applies.

The term $g_{\sigma}(y)$ in the detection probability Eq. (\ref{petb}) is of extreme importance.
This term guarantees that the detector's response to changes in its state of motion is causal and approximately local in time at macroscopic scales of observation.
To see this, we note that $g_{\sigma}(y)$ truncates contributions to the detection probability from all instants $s$ and $s'$ such that $|s-s'|$ is substantially larger than $\sigma$. Thus, at each moment of time $\tau$, the detector's response is determined solely from properties of the path at times around $\tau$ with a width of order $\sigma$. In particular, properties of the path at the asymptotic past (or future) do not affect the detector's response at time $\tau$. The response is determined solely by the properties of the path at time $\tau$, within the accuracy allowed by the detector's temporal resolution.

{\em Finite-size detectors.} The approximation Eq. (\ref{UdW}) simplifies the PDF for particle detection significantly. Eq. (\ref{UdW}) implies that the detection events are sharply localized in space, while their time localization is of order $\sigma$. A more realistic model requires the evaluation of Eq. (\ref{pet3}) without the simplifying assumption  Eq. (\ref{UdW}). Assuming that the dimensions of the detector are small in relation to the wave-length of the absorbed quanta, a meaningful approximation for the current operator is

\begin{eqnarray}
\hat{J}(x, \tau) \simeq \hat{m}  u_{\delta}[x - X^i(\tau)], \label{UdW2}
\end{eqnarray}
 where $u_{\delta}(x)$ is a function localized around $x = 0 $ with a spread equal to $\delta$. In this approximation, a detector is characterized by two scales: a spatial coarse-graining scale $\delta$ and a temporal coarse-graining scale $\sigma$. If $\sigma >> \delta$, the contribution of the temporal coarse-graining dominates and Eq. (\ref{pet5}) follows. This is the physically most relevant regime, because as we show in Sec. 3, the unless the coarse-graining parameter $\sigma$ is sufficiently large, the detection rate is strongly suppressed.

\section{Detection probabilities at different regimes}
In this section,   we evaluate  the detection probability Eq. (\ref{petb}) for different trajectories. The cases presented here are chosen in order to clarify the physical interpretation of the coarse-graining scale, and to illustrate   various techniques.  An important result is the derivation of the adiabatic approximations and of its corrections (Sec. 4.4). Furthermore, we demonstrate that the effective detection probability of fast periodic motions is constant (Sec. 4.5).

 \subsection{General expressions}
In order to evaluate the PDF Eq. (\ref{pet5}) we must first specify the function $g_{\sigma}(x)$. As a matter of fact, the
   explicit form of $g_{\sigma}$ is of little physical significance  because   $\sigma$ is a time-scale rather than a sharply defined time-parameter.  The most interesting physical predictions correspond to the asymptotic regimes with respect to $\sigma$.

We consider smearing functions  $g_{\sigma}$ that satisfy  the following  conditions
 \begin{enumerate}
 \item $g_{\sigma}(0) = 1$ and this is the single local maximum of the function.
 \item $\lim_{\sigma \rightarrow 0} g_{\sigma} = 0$; $\lim_{\sigma \rightarrow \infty} g_{\sigma} = 1$.
 \item $g_{\sigma}(y)$ drops to zero for $ y >> \sigma$.
 \item In the lower half of the imaginary plane ${\bf C}$, $g_{\sigma}(y)$ is meromorphic  and vanishes as   $y \rightarrow \infty$.
 \end{enumerate}

 Conditions 1-3 follow from the definition of $g_{\sigma}$. Condition 4 is a technical assumption that is convenient for calculational purposes.
 In what follows, we will employ the Lorentzian
 \begin{eqnarray}
 g_{\sigma}(y) = \frac{1}{\left(\frac{y}{\sigma}\right)^2 +1}, \label{lor}
 \end{eqnarray}
whenever an explicit form of $g_{\sigma}$ is needed.

 Let  $y =-i w_n(\tau)$ be the solutions of equation $\Sigma(\tau, y) = 0$ for fixed $\tau$, where $\Sigma(\tau, y)$  is given by Eq. (\ref{sigma0}). We restrict to solutions that lie in the  lower half of the imaginary plane (for $Re  w_n(\tau) >0 $), because in this half-plane $e^{-iEy}$ vanishes as $y$ goes to complex infinity. This is the reason for the  assumption 4 above.

 We assume that the solutions are simple, and we choose the index $n$ so that the solutions are indexed by their modulus: if $|w_n(t)| > |w_m(\tau)|$ then $n > m$. Then, using Cauchy's theorem, Eq.  (\ref{pet5}) becomes
 \begin{eqnarray}
 P(E, \tau) =  \frac{\alpha(E)}{2 \pi} \left(i \sum_n  \left[\left(\frac{w_n(\tau)}{\sigma}\right)^2 + 1\right]^{-1} \frac{e^{-E w_n(\tau)}}{\partial_y \Sigma[\tau, -iw_n(\tau)]} - \frac{\sigma e^{-E \sigma}}{2  \Sigma(\tau, - i \sigma) }  \right), \label{sumsol}
 \end{eqnarray}
where Eq. (\ref{lor}) was employed.

The modulus of each term in the sum over $n$ in Eq. (\ref{sumsol}) is suppressed by a factor of $(\sigma/|w_n|)^2$. Hence, poles with modulus $|w_n| >> \sigma$ contribute little to the detection probability.

 \subsection{Motion along inertial trajectories}

For any straight-line timelike path on Minkowski spacetime  we have $\Sigma(\tau, y) = y^2$. Hence, there is no pole in the lower half of the complex plane, and the detection probability Eq. (\ref{petb}) becomes
\begin{eqnarray}
 P(E, \tau)  = \frac{\alpha(E) e^{-E \sigma}}{4 \pi \sigma}.
\end{eqnarray}

Since $E \sigma >> 1$, the detection probability is strongly suppressed, as expected. It is non-zero, because any localized measurement in relativistic systems is subject to false alarms \cite{PerTer}, i.e., spurious detection events.

\subsection{Uniform proper acceleration}

Next, we consider a detector under uniform proper acceleration $a$, with a trajectory
\begin{eqnarray}
    X^{\mu}(\tau) = \frac{1}{a}(\sinh(a\tau), \cos(a\tau) - 1, 0, 0).
\end{eqnarray}
For this path, the function
\begin{eqnarray}
\Sigma(\tau, y) = \frac{4 }{a^2} \sinh^2\left(\frac{ay}{2}\right) \label{sigma1}
 \end{eqnarray}
 has double zeros at $y = - i \frac{2 \pi}{a} n$, for $n = 1, 2, \ldots$.

 The detection probability Eq. (\ref{pet5}) becomes
\begin{eqnarray}
P(E, \tau) = \frac{\alpha(E)}{2 \pi \sigma} \left\{ \frac{E}{T_a} x \left[\frac{x}{2}\left(\Phi_1(e^{-\frac{E}{T_a}}, x)- \Phi_1(e^{-\frac{E}{T_a}}, -x)\right) -1 \right] \right. \nonumber \\
\left.
+ \frac{x^2}{2}    \left(\Phi_2(e^{-\frac{E}{T_a}}, x)- \Phi_2(e^{-\frac{E}{T_a}}, -x)\right)   + \frac{\pi^2 x^2 e^{-x\frac{E}{T_a}}}{2 \sin^2(\pi x)}           \right\}, \label{pet6}
\end{eqnarray}
 where $T_a := \frac{a}{2 \pi}$ is the {\em acceleration temperature}, $x := T_a \sigma$ and $\Phi_s(z, x)$ is the Lerch-Hurwitz function \cite{GR} defined by
 \begin{eqnarray}
 \Phi_s(z, x) = \sum_{n = 0}^{\infty} \frac{z^n}{(n + x)^s}.
 \end{eqnarray}

For  $x >> 1$,   the   asymptotic expansion of Lerch-Hurwitz functions is relevant \cite{Ferreira}
\begin{eqnarray}
\Phi_1(z, x) = \frac{1}{1 -z} \frac{1}{x} + \sum_{n = 1}^{\infty} \frac{(-1)^n L_n(z)}{x^{n+1}} \\
\Phi_2(z, x) =   \frac{1}{1 -z} \frac{1}{x^2} + \sum_{n = 1}^{\infty} \frac{(-1)^n (n+1) L_n(z)}{x^{n+2}},
  \end{eqnarray}
where $L_n(z)$ is defined by the recursion relation $L_0(z) = \frac{1}{1 - z}$, $L_n(z) = z \frac{d}{dz} L_{n-1}(z)$.  We obtain 

\begin{eqnarray}
P(E, \tau) = \frac{\alpha(E) T_a}{2 \pi} \left[ \frac{E/T_a }{e^{\frac{E}{T_a}}-1} + \sum_{k = 1}^{\infty} \frac{L_{2k}(e^{-\frac{E}{T_a}}) (E/T_a) - k L_{2k-1}(e^{-\frac{E}{T_a}}) }{x^{2k}}   \right], \label{exp}
\end{eqnarray}
 where we dropped the last term in Eq. (\ref{pet6}). Eq. (\ref{exp}) is valid for energies such that $E/T_a >> x^{-1}$, which is the physically relevant regime corresponding to $E \sigma >> 1$.

 For $x \rightarrow \infty$, we recover the standard Planck-spectrum response. Including the leading correction  we obtain
 \begin{eqnarray}
 P(E, \tau)  = \frac{\alpha(E) E}{2 \pi\left(e^{\frac{E}{T_a}}-1\right)} \left[1 + \frac{1}{x^2} \frac{2 - T_a/E + e^{-\frac{E}{T_a}}(1+ T_a/E)}{\left(1 - e^{-\frac{E}{T_a}}\right)^2} + \ldots\right]. \label{corrections}
 \end{eqnarray}

 In Fig. 1, we plot the relative size $C$ of the correction terms in Eq. (\ref{exp}), as a function of $E/T_a$ and for different values of $x$. $C$ is defined as the absolute value of the second term in the parenthesis of  Eq. (\ref{exp}), modulo the first term in the parenthesis (that corresponds to the Planckian spectrum). We see that the correction to the Planckian spectrum is stronger at low energies  but in any case small for $\sigma a $ of order $10^1$.

\begin{figure}[tbp]
\includegraphics[height=7cm]{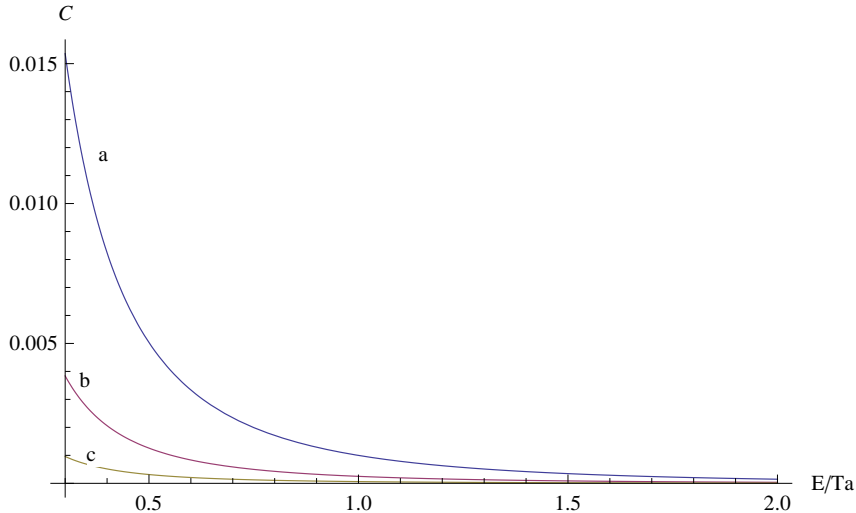} \caption{ \small The relative size $C$ of the correction terms to the Planckian spectrum in Eq. (\ref{exp}) as a function of $E/T_a$, for different values of $x$: (a) $x = 25$, (b) $x = 50$, (c) $x = 100$. The correction term is stronger at low energies; however one has to keep in mind  the condition $E/T_a >> x^{-1}$ for energy measurements.  $C$ is defined as the absolute value of the second term, modulo the first term  in the parenthesis of  Eq. (\ref{exp}).}
\end{figure}
 If $x$ is of the order of unity or smaller  then only energies $E$ such that $E >> T_a$    contribute to the detection probability. In this regime, the detection probability is exponentially suppressed for all measurable values of energy.

We conclude that only in the regime $\sigma a >> 1$ is there a significant particle detection rate, and in this regime the rate is well approximated by the Planckian spectrum expression.

 \subsection{Motion along a single axis}
 Next, we consider a general path along a single Cartesian axis in Minkowski spacetime. The four-velocity is
 \begin{eqnarray}
 \dot{X}^{\mu}(\tau) = (\cosh b(\tau), \sinh b(\tau), 0 ,0), \label{1dim}
 \end{eqnarray}
 for some function $b(\tau)$;  $\alpha(\tau) = \dot{b}(\tau) $ is the proper acceleration of the path. For this path, we find
 \begin{eqnarray}
 \Sigma(\tau, y) = \left( \int_{-y/2}^{y/2} ds e^{b(\tau + s)}\right)\left( \int_{-y/2}^{y/2} ds' e^{-b(\tau + s')}\right). \label{sigma2a}
 \end{eqnarray}

\subsubsection{Adiabatic approximation.}
The presence of the factor $g_{\sigma}$ in Eq. (\ref{petb}) implies that values of $y >> \sigma$ are suppressed. Hence, unless $b(\tau)$ varies too strongly in the scale of $\sigma$, the Taylor expansion of $b(\tau + s)$ around $b(\tau)$ in Eq. (\ref{sigma2a}) is a meaningful approximation.

The first order  in the Taylor expansion $b(\tau + s) = b(\tau) + s a_{\tau}$ corresponds to the adiabatic approximation. Writing  $a_t = \dot{b}(t)$, we obtain $\Sigma(\tau, y) = 4 \sin^2[a_{\tau}y/2]/a^2_{\tau}$. Hence,
\begin{eqnarray}
P(E, \tau)  = \frac{\alpha(E) E}{2 \pi\left(e^{\frac{2 \pi E}{a_{\tau}}}-1\right)} + O \left( \frac{1}{[\sigma a_{\tau}]^2}\right), \label{timed}
\end{eqnarray}
i.e., we obtain a Planckian spectrum with a time-dependent temperature.

\subsubsection{ Corrections to the adiabatic approximation.} Eq. (\ref{timed}) is a good approximation for paths such that $ |\dot{a}_{\tau}| \sigma << |a_{\tau}|$. In order to calculate the first corrections to Eq. (\ref{timed}), we expand $b(\tau +s)$ to second order in $s$, writing $b(\tau +s) = b(\tau) + s a_{\tau} + \frac{1}{2} s^2 \dot{a}_{\tau}$. We obtain
\begin{eqnarray}
\Sigma(\tau, y) = \frac{\pi}{\dot{a}_{\tau}} \left\{ \mbox{erfi}\left[\sqrt{\frac{\dot{a}_{\tau}}{2}}\left(\frac{a_{\tau}}{\dot{a}_{\tau}} + \frac{y}{2}\right) \right] - \mbox{erfi}\left[\sqrt{\frac{\dot{a}_{\tau}}{2}}\left(\frac{a_{\tau}}{\dot{a}_{\tau}} - \frac{y}{2}\right) \right] \right\}  \nonumber \\
\times \left\{ \mbox{erf}\left[\sqrt{\frac{\dot{a}_{\tau}}{2}}\left(\frac{a_{\tau}}{\dot{a}_{\tau}} + \frac{y}{2}\right) \right] - \mbox{erf}\left[\sqrt{\frac{\dot{a}_{\tau}}{2}}\left(\frac{a_{\tau}}{\dot{a}_{\tau}} - \frac{y}{2}\right) \right] \right\}, \label{sigma2}
\end{eqnarray}
 where $\mbox{erf}(x) =  \frac{2}{\sqrt{\pi}} \int_0^x e^{-t^2} dt$ is the error function and $\mbox{erfi}(x) = -i \mbox{erf}(ix)$. 
 
 For $|\dot{a}_{\tau}y/ a_{\tau}| << 1$, the asymptotic regime for the error function
 \begin{eqnarray}
 \mbox{erf} \simeq 1 - \frac{e^{-x^2}}{\sqrt{\pi} x} \label{erfa}
 \end{eqnarray}
 is relevant. Substituting Eq. (\ref{erfa}) in Eq. (\ref{sigma2}) we find
 \begin{eqnarray}
 \Sigma(\tau, y) = \frac{4}{a_{\tau}^2} \left[ \sinh\left(\frac{a_{\tau}y}{2}\right) - \frac{\dot{a}_{\tau} y}{2 a_{\tau}} \cosh\left(\frac{a_{\tau}y}{2}\right)\right]\left[ \sinh\left(\frac{a_{\tau}y}{2}\right) + \frac{\dot{a}_{\tau} y}{2 a_{\tau}} \cosh\left(\frac{a_{\tau}y}{2}\right)\right], \label{sigma3}
 \end{eqnarray}
 to leading order in $|\dot{a}_ty/a_t|$.

  We expect that the roots $w_n$ of Eq. (\ref{sigma3}) are small corrections to the roots obtained for $\dot{a}_t = 0$. Hence, we write $w_n = -i \frac{2 \pi}{a} n + \epsilon_n$, where $|\epsilon_n|<< \frac{2 \pi}{a} n$. The roots approximately correspond to solutions of the equation
  \begin{eqnarray}
  \tanh(a\epsilon_n/2) = \pm i n \frac{\pi \dot{a}_{\tau}}{a_{\tau}^2}. \label{tanhe}
  \end{eqnarray}

 According to Eq. (\ref{sumsol}), the only roots that contribute significantly to the detection probability are characterized by $n < N_{max}$ where $N_{max} \sim \sigma a_{\tau}/(2 \pi)$. For these values of $n$, the modulus of the right-hand-side of Eq. (\ref{tanhe}) is smaller than $| \dot{a}_{\tau} \sigma/a_{\tau}| <<1$. Hence, we can approximate the left-hand-side of Eq. (\ref{tanhe}) by setting $\tanh(x) \simeq x$. We obtain,  $\epsilon_n = \pm i n 2\pi \dot{a}_{\tau}/a_{\tau}^3$. It follows that

 \begin{eqnarray}
 w_n(\tau) = - i  \frac{2 \pi}{a_{\tau}}  \left( 1 \pm \frac{\dot{a}_{\tau}}{a_{\tau}^2} \right) n, \hspace{1cm} n = 1, 2, \ldots, N_{max}.
  \end{eqnarray}
 Variations in the acceleration results to a  split of each double root  of Eq. (\ref{sigma1}) into two single roots for Eq. (\ref{sigma3}).

 With the roots above we evaluate the PDF for particle detection
 \begin{eqnarray}
 p(E, \tau) = \frac{\alpha(E) a_{\tau}}{8 \pi^2 \delta_{\tau}} \sum_{n = 1}^{N_{max}} \frac{e^{-\frac{2 \pi E}{a_{\tau}}n} }{n} \left( e^{\frac{2 \pi \delta_{\tau}E}{a_{\tau}}n} - e^{-\frac{2 \pi \delta_{\tau}E}{a_{\tau}}n}\right), \label{pet8}
 \end{eqnarray}
where we wrote $\delta_{\tau} = \dot{a}_{\tau}/a_{\tau}^2$.

We extend the summation in Eq. (\ref{pet8}) to infinity, as terms of $n > N_{max}$ affect only  low values of energy that are physically irrelevant (they fail to satisfy the condition $E \sigma >> 1$). We obtain
\begin{eqnarray}
p(E, \tau) = \frac{\alpha(E) a_{\tau}}{8 \pi^2 \delta_{\tau}} \left[ g_1\left( \frac{2 \pi}{a_{\tau}} ( 1 - \delta_{\tau})\right)-  g_1\left( \frac{2 \pi}{a_{\tau}} ( 1 + \delta_{\tau})\right) \right], \label{petg}
\end{eqnarray}
where $g_1(x) = \sum_{n=1}^{\infty}\frac{e^{-nx}}{n}$.

Eq. (\ref{petg}) applies to all energies $E$, such that $E\sigma >> 1$. If we exclude very high energies of order $a_{\tau}/\delta_{\tau}$  we can expand the function $g_1$ with respect to $\delta_{\tau}$. Thus, we obtain the leading correction to the Planckian spectrum due to small variations in the acceleration
\begin{eqnarray}
P(E, \tau)  = \frac{\alpha(E) E}{2 \pi\left(e^{\frac{2 \pi E}{a_{\tau}}}-1\right)} \left[1 + \frac{2 \pi^2 \dot{a}_{\tau}^2 E^2}{3 a_{\tau}^6} \frac{  1 +  e^{-\frac{2 \pi E}{a_{\tau}}}}{\left(1 - e^{-\frac{2 \pi E}{a_{\tau}}}\right)^2} \right]+ O \left( \frac{1}{[\sigma a_{\tau}]^2} , \frac{\dot{a}_{\tau}\sigma}{a_{\tau}}\right). \label{pet11}
\end{eqnarray}

Note that Eq. (\ref{pet11})   applies in the regime where both $\sigma a_{\tau} >> 1$ and $|\dot{a}_{\tau} \sigma/a_{\tau}| << 1$. The introduction of $\sigma$ is essential for the separation of timescales leading to the probability density Eq. (\ref{pet11}), even though $\sigma$ does not explicitly appear in the equation.

\subsubsection{Non-relativistic limit}
In the non-relativistic regime, $b(\tau)$ coincides with the non-relativistic velocity $v_{\tau}$ and satisfies $|v(\tau)|<<1$. We write the exponentials of Eq. (\ref{sigma2a}) as $e^{\pm b} \simeq 1 \pm v$ and we Taylor-expand $b(\tau +s)$ around $\tau$. If the coarse-graining time-scale $\sigma$ is sufficiently large then  a finite number of terms in the Taylor expansion suffices. Thus, $\Sigma(\tau, y)$ becomes a product of two polynomials, and the full set of its roots can be easily determined.

To the lowest-non trivial order,
\begin{eqnarray}
\Sigma(\tau, y) = y^2 \left( 1 + v_{\tau} + \frac{\dot{a}_{\tau}}{24} y^2 \right) \left( 1 - v_{\tau} - \frac{\dot{a}_{\tau}}{24} y^2 \right). \label{sigma6}
\end{eqnarray}
 The only root of $\Sigma(\tau, y)$ in the lower half of the complex plane is $w_1 = -2  i \sqrt{6(1 + v_{\tau})/|\dot{a}_{\tau}|} \simeq -2  i \sqrt{6/|\dot{a}_{\tau}|}$. However, the Taylor-expansion of $v(\tau + s)$  preserve the condition $|v_{\tau}| << 1$ only if
 $|\dot{a}_{\tau}|\sigma^2 << 1$. This implies that $|w_1|>> \sigma$, and hence, that the contribution of $w_1$ to the detection probability is suppressed. Thus, the detection probability effectively vanishes.

 The conclusion above is not affected when keeping higher order terms in the Taylor expansion of $v(\tau+s)$; the relevant roots of Eq. (\ref{sigma6}) are much larger in norm than $\sigma$ because the velocity is restricted in the non-relativistic regime. Hence, particle detection vanishes in the non-relativistic regime.  The only possible exceptions  are paths that are characterized by rapid variations of their velocity at scales much smaller than $\sigma$. For such paths, any finite number of terms in the Taylor expansion of $v(\tau + s)$ provides an inadequate approximation. Such is the case of rapid oscillations, which we examine next.

\subsection{Periodic motions}

In what follows, we show that the effective detection rate for periodic motions is constant, provided that the period $T$ is much shorter than the coarse-graining time-scale $\sigma$.

\subsubsection{Time-averaging}

Consider a spacetime path of the Unruh-DeWitt detector with strong variations at time-scales that are much shorter than $\sigma$. The details of such variations are unobservable; detection events are localized at a coarser time-scale. The physical probability density involves a smearing of $\tau$ in  $P(E, \tau)$ at a scale $\sigma$, as given in Eq. (\ref{conv}). We must recall that the smeared form Eq. (\ref{conv}) is derived from first principles  and that the de-convoluted PDF is a convenient approximation.

The use of the smeared probability density $\bar{P}(E, \tau)$  is equivalent to the substitution of the term $1/ \Sigma(\tau, y)$ in Eq. (\ref{pet5})   with a term $1 / \bar{\Sigma}(\tau, y) := \langle 1/ \Sigma(\tau, y)\rangle_{\sigma}$, where

\begin{eqnarray}
\langle A(\tau) \rangle_{\sigma} = \int d \tau' f_{\sigma}(\tau - \tau') A(\tau'),
\end{eqnarray}
defines that average of any function $A(\tau)$ with the probability density $f_{\sigma}$ of Eq. (\ref{conv}).

To a first approximation,
\begin{eqnarray}
\bar{\Sigma}(\tau, y) \simeq  \langle  \Sigma(\tau, y)\rangle_{\sigma} = \int_{-y/2}^{y/2}  ds \int_{-y/2}^{y/2}  ds' \langle \dot{X}^{\mu}(\tau + s) \dot{X}_{\mu}(\tau + s')\rangle_{\sigma}
\end{eqnarray}

If the   detector's trajectory is exactly periodic with period $T$, and $T << \sigma$, then the $\tau$ dependence drops from $1/\Sigma$ after averaging. This implies that we can substitute averaging with respect to $f_{\sigma}$ with averaging over the period, i.e.,

\begin{eqnarray}
\langle A(\tau) \rangle_{\sigma} = \frac{1}{T} \int_0^T d\tau A(\tau) = \langle A\rangle_T
\end{eqnarray}
for any variable $A(\tau)$. Hence, the smeared PDF $\bar{P}(E, \tau)$  becomes time-independent
\begin{eqnarray}
\bar{P}(E) = - \frac{\alpha(E)}{4 \pi^2} \int_{-i\epsilon - \infty}^{-i \epsilon + \infty} dy \frac{ g_{\sigma}(y) e^{-i Ey}}{\bar{\Sigma}(y)}, \label{pet10}
\end{eqnarray}
where
\begin{eqnarray}
\bar{\Sigma}(y) = \int_{-y/2}^{y/2}  ds \int_{-y/2}^{y/2}  ds' \langle L_s[\dot{X}^{\mu}] L_{s'}[\dot{X}_{\mu}]\rangle_{T} \label{sls}
\end{eqnarray}
and  $L_s$ stands for the translation operator $L_s[F](\tau) = F(\tau +s)$.

\subsubsection{Non-relativistic systems}

 Eq. (\ref{pet10}) applies to any periodic motion in Minkowski spacetime. It simplifies significantly for non-relativistic systems, where $\dot{X}^0(\tau) \simeq 1$ and $|\dot{X}^i(\tau)| << 1$. To see this, we decompose $X^i(\tau)$ in its Fourier modes

\begin{eqnarray}
X^i(\tau) = \frac{X_0^i}{2} + \sum_{n = 1}^{\infty} c^i_n \cos ( n \omega_0 \tau + \phi^i_n), \label{xin}
\end{eqnarray}
where $\omega_0 = 2 \pi/T$. Then, Eq. (\ref{sls}) becomes

\begin{eqnarray}
\bar{\Sigma}(y) =   y^2 - {\cal F} (y),
\end{eqnarray}
where
\begin{eqnarray}
{\cal F} (y) =   \sum_{n=1}^{\infty} \overrightarrow{c}_n^2  \sin^2\left(\frac{n \omega_0 y}{2} \right).
\end{eqnarray}
is a periodic function of period $T$.

Thus, the calculation of the detection probability for an oscillator in periodic non-relativistic motion reduces to finding the roots of the equation $z^2 - {\cal F}(z) = 0$, in the lower half of the imaginary plane with modulus of order $\sigma$ or smaller.

For harmonic motion of frequency $\omega$, $X^i(\tau) = x^i_0 \cos(\omega \tau + \phi^i)$. Hence,
\begin{eqnarray}
{\cal F}(y) = \overrightarrow{x}^2_0 \sin^2 \left(\frac{\omega y}{2}\right)
\end{eqnarray}
and the detection PDF coincides with that for uniform circular motion  at angular frequency $\omega$, as it has been studied in Refs. \cite{LePf, Letaw}.

%The calculation of the detection probability requires the roots of the equation $z = \pm v_0 \sin z$ in the lower half of the imaginary plane, where $v_0 = \omega |\overrightarrow{x}_0| <<1$ is the maximum oscillation velocity. Then $\bar{\Sigma}(y)$ coincides with the expression for circular motion---see Sec. 3.7.

\subsubsection{Relativistic harmonic oscillation}
We calculate   the time-averaged detection probability for relativistic harmonic motion in one dimension, a case that is of interest for experimental reasons. We consider a path characterized by the spatial oscillatory motion $X^1(\tau) = x_0 \sin(\omega \tau)$. This path is of the form Eq. (\ref{1dim}), where  $b(\tau) = \sinh^{-1}[v_0 \sin(\omega \tau)]$ and $v_0 = \omega x_0 < 1$.

We expand $e^{\pm b(\tau)}$ as a Fourier series
\begin{eqnarray}
e^{\pm b(\tau)} = \pm v_0 \cos \left( \omega \tau \right) + \sqrt{1 + v_0^2 \cos^2\left(\omega \tau \right)} = \pm v_0 \cos \left( \omega \tau \right) + \frac{c_0(v_0)}{2}  + \sum_{k = 1}^{\infty} c_k(v_0) \cos \left(2k \omega \tau \right),
\end{eqnarray}
where
\begin{eqnarray}
c_k(v_0) = 2 \sum_{n = k}^{\infty} (-1)^{n-1} \frac{1 \cdot 3 \cdot 5 \cdot 7\cdot \ldots \cdot (2n - 1) }{(n-k)!(n+k)!n!} \left( \frac{v_0^2}{8}\right)^{n}, \hspace{0.5cm} k = 0, 1, 2, \ldots.
\end{eqnarray}
Then, we find
\begin{eqnarray}
\langle L_s[\dot{X}^{\mu}] L_{s'}[\dot{X}_{\mu}]\rangle_{T} = \langle L_s[e^b] L_{s'}[e^{-b}]\rangle_{T} = \frac{c_0^2}{4} - \frac{v_0^2}{2} \cos\left[\omega(s-s')\right] + \frac{1}{2} \sum_{k = 1}^{\infty} c_k^2 \cos \left[ 2 k \omega (s -s') \right],
\end{eqnarray}
and we compute the time-averaged proper distance
\begin{eqnarray}
\bar{\Sigma}(y) = \frac{c_0^2}{4} y^2 - \frac{2 v_0^2}{\omega^2} \sin^2\left(\frac{\omega y}{2} \right) + \sum_{k = 1}^{\infty} \frac{2 c_k^2}{\omega^2 k^2} \sin^2\left(k \omega y \right). \label{sigma13}
\end{eqnarray}
The time-averaged detection probability $\bar{P}(E)$ associated to Eq. (\ref{sigma13}) is computed numerically. Fig.2 is a logarithmic plot of $\bar{P}(E)/\alpha(E)$,  for different values of $v_0$ in the relativistic regime.

\begin{figure}[tbp]
\includegraphics[height=7cm]{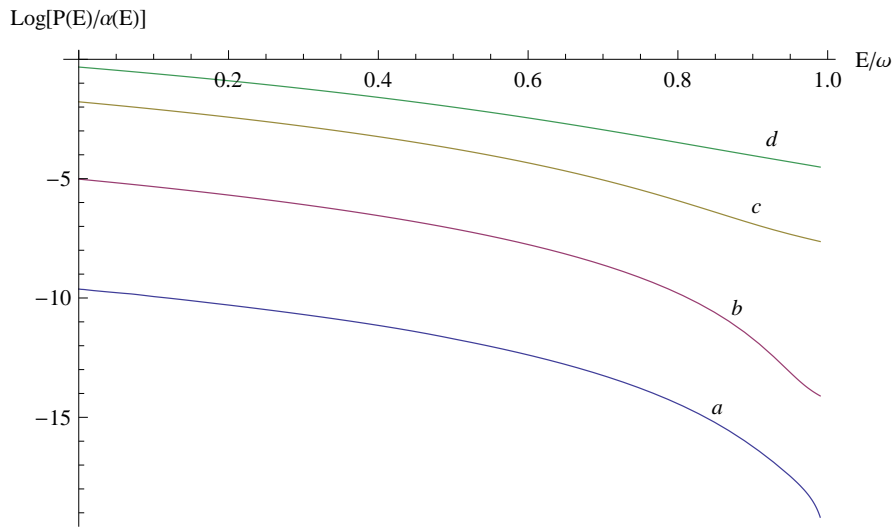} \caption{ \small The PDF $\bar{P}(E)$ divided by $\alpha(E)$, Eq. (\ref{ae}), as a function of $E/\omega$ for a detector undergoing an harmonic oscillation at frequency $\omega$. The scale on the vertical axis is logarithmic, as the detection probability varies strongly with the maximum velocity $v_0$ of the oscillation; (a) $v_0 = 0.01$, (b) $v_0 = 0.1$, (c) $v_0 = 0.5$, and (d) $v_0= 0.99$.  }
\end{figure}

\subsection{ Averaging over fast motions}
We saw that the averaged detection PDF is constant for periodic motions
with period $T$  much smaller than the coarse-graining scale $\sigma$. We expect that the detection probability will vary slowly in time in quasi-periodic motions, i.e., in motions that are almost periodic at a scale $T$ but non-periodic when examined at a larger scale.

An example of such a quasiperiodic motion is an harmonic motion $X^i(\tau) = x^i_0 \cos(\omega_{\tau} \tau)$, where $\omega(\tau)$ is a function of time that varies at time-scales much large than $\omega^{-1}$. In particular, if $\dot{\omega} \sigma/\omega << 1$, by continuity we expect that  to leading order  Eq. (\ref{pet10}) applies for the detection probability, with a time-dependence due to the variation of $\omega_{\tau}$.

In general, time-averaging allows for the elimination of motions at scales much smaller than $\sigma$, usually resulting in a simpler description at the time-scale of observation. To see this, we consider   a  path of the form Eq. (\ref{1dim})  with
\begin{eqnarray}
b(\tau) = a_0 \tau + \frac{a_1}{\omega} \sin (\omega \tau)
\end{eqnarray}
where $a_1 << a_0$ and $a_1 << \omega$. This path is characterized by constant acceleration $a_0$ modulated by small but rapid oscillations at frequency $\omega$. It is not well described by Eq. (\ref{petg}), because $\dot{a}_{\tau}/a_{\tau}^2 \sim a_1 \omega/a_0^2$ is not necessarily much smaller than unity. For this path,
\begin{eqnarray}
\bar{\Sigma}(y) = \int_{-y/2}^{y/2}  ds \int_{-y/2}^{y/2}  ds' e^{a_0(s -s') }\langle e^{\frac{a_1}{\omega} [\cos(\omega \tau + s) - \cos[\omega \tau + s')]} \rangle_T. \label{sig6}
\end{eqnarray}
Since $a_1/\omega << 1$, we expand the exponential in Eq. (\ref{sig6}) to obtain
\begin{eqnarray}
\bar{\Sigma}(y)  = \frac{4}{a_0^2} \sinh^2\left(\frac{a_0y}{2} \right) - \frac{a_1^2}{\omega^2}  \frac{(a_0^2 - \omega^2) [\cosh(a_0y) \cos(\omega y) - 1]+ 2 a_0 \omega \sinh(a_0y) \sin(\omega y)}{(a^2_0 + \omega^2)^2}
\end{eqnarray}
In the regime  $\omega >> a_0$, the roots $w_n(\tau)$ of $\bar{\Sigma}(y)$ are obtained perturbatively as in Sec. 3.4. We find
\begin{eqnarray}
w_n(\tau) = -i\frac{2 \pi }{a_0} \left(1 \pm \frac{\sqrt{2}a_1}{\omega} \right)n, \hspace{1cm} n < N_{max} << \frac{\omega}{a_0} \log\left(\frac{\omega}{a_1}\right).
\end{eqnarray}
Following the same methodology as the one leading to  Eq. (\ref{pet11}), we compute the averaged detection probability
\begin{eqnarray}
\bar{P}(E) = \frac{\alpha(E) E}{2 \pi\left(e^{\frac{2 \pi E}{a_0}}-1\right)} \left[1 + \frac{4 \pi^2 a_1^2 E^2}{3 a_0^2 \omega^2} \frac{  1 +  e^{-\frac{2 \pi E}{a_0}}}{\left(1 - e^{-\frac{2 \pi E}{a_0}}\right)^2} \right], \label{pet13}
\end{eqnarray}
for the corrections to the Planckian spectrum due to rapid oscillations.

\subsection{General spacetime trajectories}

The common feature of all calculations so far has been that the detection probability is non-zero  only for trajectories characterized by at least one  time-scale   $T$ that is much smaller than $\sigma$.  This result is physically reasonable. The emergence of macroscopic records at a scale of $\sigma$ implies that quantum processes at time-scales larger than $\sigma$ are decohered. In contrast, quantum processes   at time-scales much smaller than $\sigma$ remain unaffected. Therefore, it is essential that at least one of the characteristic time-scales of the detector's motion be much smaller than $\sigma$.

A generic path $X^{\mu}(\tau)$ in Minkowski spacetime is characterized by three Lorentz-invariant functions of proper time, with values having  dimension of inverse time: the acceleration $a(\tau)$, the torsion ${\cal T}(\tau)$ and the hypertorsion $ \upsilon(\tau)$. These are defined as follows
\begin{eqnarray}
a(\tau) :&=& \sqrt{-a_{\mu}a^{\mu}}, \\
{\cal T}(\tau) :&=& \frac{\sqrt{a^4 - \dot{a}^2-\dot{a}_{\mu} \dot{a}^{\mu}}}{a},\\
\upsilon(\tau) :&=& \frac{\epsilon_{\mu \nu \rho \sigma}\dot{X}^{\mu} a^{\nu} \dot{a}^{\rho} \ddot{a}^\sigma }{a^3 {\cal T}^2},
\end{eqnarray}
where $a^{\mu} = \ddot{X}^{\mu}$.

Our previous analysis suggests that the detection probability will be significant  whenever any of the following conditions hold: $ a \sigma >> 1$, ${\cal T} \sigma >> 1$, or $\upsilon \sigma >> 1$.  The converse does not hold necessarily; a generic path is characterized by other parameters arising from higher order derivatives of the functions $a, {\cal T}$ and $\upsilon$.

Of particular importance are the so-called {\em stationary paths}, i.e., paths characterized by a proper distance $\Sigma(\tau, y)$ that is independent of $\tau$. These paths correspond to constant values of $a, {\cal T}$ and $\upsilon$. They separate naturally into six classes. We have already encountered two classes corresponding to straight line motion ($ a = {\cal T} = \upsilon = 0$) and to constant linear acceleration ($a \neq 0$, $ {\cal T} = \upsilon = 0$). A third class that corresponds to circular motion ($\upsilon = 0$, $|a| < |{\cal T}|$) falls also under the category of periodic motions, described in Sec. 3.5. A representative path of the latter class is
\begin{eqnarray}
X^{\mu}(\tau) = \frac{1}{\omega^2} ({\cal T} \omega \tau, a \cos( \omega \tau), a \sin (\omega \tau)),
\end{eqnarray}
where $\omega = \sqrt{{\cal T}^2 - a^2}$ is the angular frequency of the circular motion. For this trajectory,  
\begin{eqnarray}
\Sigma(\tau, y) = \frac{{\cal T}^2}{\omega^2} [y^2 - \frac{a^2}{{\cal T}^2 \omega^2} \sin^2( \omega y)]
\end{eqnarray}
is formally similar to the  expression for non-relativistic harmonic motion in Sec. 3.5.

  Two other classes that correspond to spatially unbounded trajectories also cannot be analytically evaluated. For a numerical evaluation that corresponds to the regime $\sigma \rightarrow \infty$ of our formalism see Ref.  \cite{Letaw}.
   
   The last case corresponds to $a = {\cal T}, \upsilon = 0$ and it corresponds to a peculiar cusped motion, as described by the representative path
\begin{eqnarray}
X^{\mu}(\tau) = (\tau + \frac{1}{6}a^2 \tau^3, \frac{1}{2} a \tau^2, \frac{1}{6} a^2 \tau^3, 0),
\end{eqnarray}
for which $\Sigma(\tau, y) = y^2 \left( 1 + \frac{a^2}{12} y^2 \right)$. Then Eq. (\ref{pet5}) becomes
\begin{eqnarray}
P(E, \tau) = \frac{\alpha(E) a}{8 \sqrt{3} \pi \left(1 - \frac{12}{(\sigma a)^2}\right)} \left[ e^{-\frac{2\sqrt{3}E}{a}} - \frac{24\sqrt{3}}{(\sigma a)^3} e^{-\sigma E}\right]. \label{cusp}
\end{eqnarray}
From Eq. (\ref{cusp}) we readily verify that the detection probability is suppressed unless $\sigma a >> 1$ and that the corrections to the asymptotic value at $\sigma \rightarrow \infty$ are of order $(\sigma a)^{-2}$.

For paths $X^{\mu}(\tau)$ characterized by slow changes to the invariants $a, {\cal T}$ and $\upsilon$ (for example $|\dot{a} \sigma/a| <<1$), we expect that the adiabatic approximation will be applicable, and that the corrections to the adiabatic approximation will be of order $\dot{a}/a^2, \dot{{\cal T}}/{\cal T}^2$ and $\dot{\upsilon}/\upsilon^2$---see Sec. 3.4.2.

 \section{Conclusions}

In this article, we constructed the particle-detection probability for macroscopic detectors moving along general trajectories in Minkowski spacetime. The detectors are macroscopic in the sense that a particle detection is expressed in terms of definite records of observation. We describe the detection process in terms of PDF for the detection time. The derivation of this PDF is probabilistically sound and
  takes into account the irreversibility due to the emergence of measurement records in a detector.

  The resulting PDF is causal and local in time at macroscopic time-scales. We found that a key role is played by the time-scale $\sigma$ of the temporal coarse-graining necessary for the creation of a macroscopic record. Detectors moving along paths with characteristic time-scales   of order $\sigma$ or larger do not click. This behavior is physically sensible:    particle creation  depends strongly on the coherence properties of the quantum vacuum and the emergence of records at a time-scale of $\sigma$ destroys the coherence of all processes at larger timescales.

For paths characterized by multiple time-scales,   $\sigma$ provides a scale by which to distinguish between slow and fast variables. Slow variables can be treated with an adiabatic approximation, and corrections to the adiabatic approximation can be systematically calculated.   Moreover,  the PDF  Eq. (\ref{petb}) can be averaged over all fast processes. We showed that for {\em any} periodic motion of period $T << \sigma$, the effective detection probability is time-independent.

We believe that our results are important for the conceptual clarification of the role of the detector in the Unruh effect. In particular, the consideration of macroscopic detectors is one step towards a thermodynamic description of the Unruh effect, because the recorded energy can be interpreted as heat. Indeed, the detectors presented here can be viewed as a quantum models for a calorimeter.

In order to explain this point, we note that an interpretation of the Unruh temperature as a thermodynamic temperature faces the problem of explaining what is the physical system to which this temperature is attributed. The analogy with the Hawking temperature does not hold here  because the Hawking temperature is attributed to a black hole.
 There are several proposals that the Unruh temperature can be interpreted as a temperature of the spacetime \cite{thermgrav}. In this viewpoint,   general relativity is viewed as an emergent thermodynamic description of an underlying theory.

     While unrelated  to such proposals, our results strengthen the idea that the Unruh temperature can be interpreted thermodynamically. First, we establish the robustness of the Unruh effect, in the sense that slow changes in a detector's acceleration correspond to slow changes in the temperature of the Planckian spectrum. Second, our formulation in terms of the macroscopic response of  detectors and the emphasis on the role of coarse-graining is structurally compatible with the foundations of statistical mechanics.

{\em Experimental tests.} Starting with Bell's and Leinaas' proposal of an Unruh effect interpretation of spin depolarization of electrons in circular motion \cite{BeLe, Uncir}, several proposals have been formulated for measuring the Unruh effect,  or more generally particle detection due to non-inertial motion \cite{rogers, MaVa, ChTa, skbfc, SSH, MFM}. Such experiments are mostly concerned with microscopic systems rather than macroscopic detectors, as described here.

In order to identify what kind of physical system could play the role of a macroscopic detector that effectively  records particle creation, we recall that the   coarse-graining time-scale $\sigma$   must be significantly larger than $(\Delta E)^{-1}$, where $\Delta E$ is the uncertainty in the energy of the pointer variable. Supposing that the pointer variable corresponds to a collective excitation of $N$-particles, we estimate $\Delta E$ as the energy fluctuations in the canonical ensemble for an $N$-particle system at temperature $T$. Then, $\Delta E = k(T) \sqrt{N}$, where $k(T)$ is a function that vanishes as $T \rightarrow 0$. Hence,   $\sigma $ decreases with increasing $N$ as $N^{-1/2}$, but increases as $T$ approaches zero. The ideal  detector should have the minimum number of particles consistent with the formation of stable records of detection, and it should be prepared at temperatures close to zero.

However, our method also applies when the detector consists of a microscopic probe (such as an atom in a trap) {\em and} a macroscopic apparatus   (such as a photodetector) that detects the photons emitted by the atom, provided that atom and detector follow the same trajectory.
 Proposed experiments involve an accelerated microscopic probe together with detectors that are  at rest in the laboratory frame, so Eq. (\ref{petb}) does not apply. However,  a theoretical model that involves only the coupling of the probe to the field misses the fact that the records of observation corresponds to photon emitted by the probe after it has been excited. 
 
 The non-equilibrium dynamics of an accelerating probe coupled to the field is, in general, non-Markovian \cite{Hual}, so there is no obvious relation between the state of the probe and the recorded detection rate. Furthermore, as shown here, the temporal coarse-graining inherent in the detection process may destroy the particle detection signal. 
 
 For the reasons above, we believe that a complete theoretical analysis of such experiments requires
    not only the study of the non-equilibrium dynamics of the probe coupled to the field \cite{Hual, Hual2}, but also a precise modeling of the detection process along the lines presented here.

Finally, we note that our treatment of macroscopic detectors has the additional benefit that it allows for the explicit construction of multi-time correlation functions \cite{AnSav11}  known as {\em coherence} in quantum optics.  The consideration of the detection correlations for multiple detectors, along different spacetime trajectories,   may lead to a new method for verifying phenomena of particle creation due to non-inertial  motion.


\begin{thebibliography}{c}


\bibitem{Unruh} W. G. Unruh, Phys. Rev. D 14, 870 (1976).

\bibitem{Boyer} H. Boyer, Phys. Rev. D21, 2137 (1980).

\bibitem{Dewitt} B. S. DeWitt, in {\em General Relativity: An Einstein Centenary Survey}, ed. by S. W. Hawking and W. Israel (Cambridge
University Press, Cambridge, 1979), p. 680.



\bibitem{AnSav11} C. Anastopoulos and N. Savvidou,  J. Math. Phys. 53, 012107 (2012).


\bibitem{AnSav12} C. Anastopoulos and N. Savvidou, Phys. Rev. A86, 012111 (2012).


\bibitem{Sav99} K. Savvidou,   J. Math. Phys. 40, 5657 (1999).

\bibitem{Sav10} N. Savvidou,   in {\em Approaches to Quantum Gravity}, edited by D. Oriti (Cambridge University Press, Cambridge 2010).





 \bibitem{AnSav06} C. Anastopoulos and N. Savvidou,   J. Math. Phys. 47, 122106 (2006).


\bibitem{AnSav08} C. Anastopoulos and N. Savvidou,   J. Math. Phys. 49, 022101 (2008).

\bibitem{AnSav13} C. Anastopoulos and N. Savvidou, Ann. Phys. 336, 281 (2013).




 \bibitem{An08} C. Anastopoulos,   J. Math. Phys. 49, 022103 (2008).


\bibitem{WM} See, for example, D. F. Walls and G. J. Milburn, {\em Quantum Optics} (Springer, 2010).




\bibitem{SvSv} B. F. Svaiter and N. F. Svaiter,   Phys. Rev. D46,
5267 (1992).

\bibitem{Hig93} A. Higuchi, G. E. A. Matsas, and C. B. Peres,  Phys. Rev. D 48,
3731 (1993).

\bibitem{SPad96} L. Sriramkumar and T. Padmanabhan,   Class. Quant. Grav. 13, 2061 (1996).


\bibitem{Schlicht}S. Schlicht,    Class. Quant. Grav. 21, 4647 (2004).

\bibitem{Langlois} P. Langlois, Ann. Phys. (N.Y.) 321, 2027 (2006).

\bibitem{LoukoSatz}J. Louko and A. Satz, Class. Quant. Grav. 23, 6321 (2006).

\bibitem{Milgrom} N. Obadia and M. Milgrom, Phys. Rev. D 75, 065006 (2007).

\bibitem{HuLin} Y-S Lin and B. L. Hu, Phys. Rev. D76, 064008 (2007).

\bibitem{BoRo} N. Bohr and L. Rosenfeld, Mat. Fys. Medd. Dan. Vid. Selsk.
12,  8 (1933); Engl. transl.  in
{\em Selected Papers of L\'eon Rosenfeld},
eds R S Cohen and J
Stachel (Dordrecht: Reidel, 1979) p. 357, reprinted in
{\em Quantum Theory and
Measurement}
eds J A Wheeler and W H Zurek (Princeton, New Jersey: Princeto
n U.P., 1983 ) p.
479.



\bibitem{Omn} R. Omn\'es, {\em The Interpretation of Quantum Mechanics}, (Princeton University Press, 1994).


\bibitem{Omn2}  R. Omn\'es,	{\em Understanding Quantum Mechanics} (Princeton University Press, 1999).
\bibitem{Gri} R. B. Griffiths, {\em 	Consistent Quantum Theory} (Cambridge University Press, 2003).


\bibitem{GeHa93} M. Gell-Mann and J.  B.  Hartle,  in {\em Complexity, Entropy and the Physics of Information}, edited by W. Zurek   (Addison Wesley, Reading, 1990) ; Phys. Rev.   D47, 3345 (1993).

\bibitem{hartlelo} J. B. Hartle, "Spacetime quantum mechanics and the quantum mechanics of spacetime" in {\em Proceedings on the 1992 Les Houches School,Gravitation and Quantization}  (1993).



\bibitem{Tagaki} S. Takagi,
Prog. Th. Phys. Supp. 88, 1 (1986).



\bibitem{PerTer} A. Peres and D. R. Terno,   Rev. Mod. Phys. 76, 93 (2004).



\bibitem{GR} I. S. Gradshteyn and I. M. Ryzhik, {\em Tables of Integrals, Series, and Products} (4th ed.,New York: Academic Press, 1960).

\bibitem{Ferreira} C. Ferreira and J. L. Lopez, J. Math. Anal. Appl. 298, 210 (2004).


\bibitem{LePf} R. Letaw and J. D. Pfautsch, Phys. Rev. D22, 1345 (1980).


\bibitem{Letaw} J. R. Letaw, Phys. Rev. D23, 1709 (1981).


\bibitem{thermgrav} See, for example, T. Jacobson, Phys. Rev. Lett. 75, 1260 (1995); T. Padmanabhan, Gen. Rel. Grav. 34, 2029 (2002); Phys. Rept. 406, 49 (2005); T. Padmanabhan, Rep. Prog. Phys. 73, 046901 (2010); E.  P. Verlide, JHEP 1104, 029 (2011).


\bibitem{BeLe} J. S. Bell and J. M. Leinaas, Nucl. Phys. B
212
, 131 (1983); J. S. Bell and J. M. Leinaas, Nucl. Phys. B
284
488 (1987).

\bibitem{Uncir} W. G. Unruh, in {\em Monterey Workshop on Quantum Aspects of Beam Physics}, edited by P. Chen (World Scientific,
Singapore, 1998); Phys. Rep.
307
163 (1998).


\bibitem{rogers}J. Rogers, Phys. Rev. Lett.
61, 2113 (1988).

\bibitem{MaVa} G. E. A. Matsas and D. A. T. Vanzella, Phys.
Rev. D
59, 094004 (1999).


\bibitem{ChTa} P. Chen and T. Tajima, Phys. Rev. Lett.
83
, 256 (1999).

\bibitem{skbfc} M. O. Scully, V. V. Kocharovsky, A. Belyanin, E. Fry, and F. Capasso,
Phys. Rev. Lett.
91, 243004 (2003);93, 129302 (2004); B. L. Hu and A. Roura, Phys. Rev. Lett.
93, 129301 (2004).

\bibitem{SSH}R.
Sch\"utzhold, G. Schaller, and D. Habs, Phys. Rev. Lett.
97
, 121302 (2006).




\bibitem{MFM} E. Martin-Martinez, I. Fuentes, and R. B. Mann, Phys. Rev. Lett.
107
, 131301 (2011).


\bibitem{Hual} J. Doukas, S. Y. Lin, B. L. Hu and R. B. Mann, JHEP 11, 119 (2013).

\bibitem{Hual2}  D. C. M. Ostapchuk, S.Y. Lin, R. B. Mann, and B. L. Hu, JHEP
1207
, 072 (2012)







%\bibitem{Kij} J. Kijowski, Rep. Math. Phys. 6, 361 (1974).


\end{thebibliography}
\end{document}